\documentclass{aa}
\usepackage[varg]{txfonts}
\usepackage[utf8]{inputenc}
\usepackage[english]{babel}
\usepackage{natbib}
\usepackage{graphicx}
\usepackage{amstext}
\usepackage{multirow}

\newcommand{\degree}{^\circ}

\begin{document}
\title{A method to discriminate solar and antisolar differential
  rotation in high-precision light curves}

\author{Timo Reinhold\inst{1}, Rainer Arlt\inst{2}}

\offprints{T. Reinhold, \\ \email{reinhold@astro.physik.uni-goettingen.de} }

\institute{Institut für Astrophysik, Universität Göttingen, Friedrich-Hund-Platz 1, 37077
Göttingen, Germany \and 
Leibniz-Institut für Astrophysik Potsdam (AIP), An der Sternwarte 16, 14482 Potsdam,
Germany}

\date{Received day month year / Accepted day month year}

\abstract
{
Surface differential rotation (DR) is one major ingredient of the magnetic field 
generation process in the Sun and likely in other stars. The term \textit{solar-like}
differential rotation describes the observation that solar equatorial regions rotate
faster than polar ones. The opposite effect of polar regions rotating faster than
equatorial ones (termed as \textit{antisolar} DR) has only been observed in a few stars,
although there is evidence from theoretical dynamo models.
}
{
We present a new method to detect the \textit{sign} of DR (i.e. solar-like or antisolar
DR) by analyzing long-term high-precision light curves with the Lomb-Scargle periodogram.
}
{
We compute the Lomb-Scargle periodogram and identify a set of significant periods $P_k$,
which we associate with active regions located at different latitudes on the the stellar
surface. If detectable, the first harmonics ($P_k'$) of these periods were identified to
compute their peak-height-ratios $r_k:=h(P_k')/h(P_k)$. Spots rotating at lower latitudes
generate less sine-shaped light curves, which requires additional power in the harmonics,
and results in larger ratios $r_k$. Comparing different ratios $r_k$ and the associated
periods $P_k$ yields information about the spot latitudes, and reveals the sign of DR.
}
{
We tested our method on different sets of synthetic light curves all exhibiting solar-like
DR. The number of cases where our method detects antisolar DR is the false-positive rate
of our method. Depending on the set of light curves, the noise level, the required minimum
peak separation, and the presence or absence of spot evolution, our method fails to detect
the correct sign in at most 20\,\%. We applied our method to 50 Kepler G stars and found
21--34 stars with solar-like DR and 5--10 stars with antisolar DR, depending on the
minimum peak separation.
}
{
The method is capable of determining the sign of DR in a statistical way with a rather low
false-positive rate. Applying our method to real data might suggest that -- within the
uncertainties -- antisolar DR was detected in 5--10 Kepler stars.
}


\titlerunning{A method to discriminate solar and antisolar differential
  rotation in high-precision light curves}
\authorrunning{Timo Reinhold \& Rainer Arlt} 
\maketitle

\section{Introduction}\label{intro}
The differential rotation (DR) of the solar surface (and its interior) is a crucial part
of the generation of the Sun's magnetic field and the solar dynamo. The differential
rotation winds up magnetic field lines, which can be considered as frozen in the plasma,
transferring an initial poloidal field into a toroidal one. Speaking of \textit{solar} or
\textit{solar-like} differential rotation in the following we always think of
latitudinal-dependent surface rotation rates, with equatorial regions rotating faster 
than polar ones. This latitudinal shear is usually described by the equation
\begin{equation}\label{drotlaw}
 \Omega(\theta) = \Omega_{\rm eq}(1-\alpha\sin^2\theta),
\end{equation}
where $\Omega_{\rm eq}$ denotes the equatorial angular velocity, $\theta$ the latitude, 
and $\alpha$ a dimensionless parameter describing the amount of surface shear. The Sun 
has a relative shear of $\alpha_\odot=0.2$. According to eq.~\ref{drotlaw} DR in stars
other than the Sun is usually termed \textit{solar-like} when $\alpha > 0$, whereas the
opposite effect, $\alpha < 0$, is called \textit{antisolar}; the case $\alpha=0$
describes rigid (surface) rotation. For other purposes DR might also be expressed in terms
of the absolute surface shear 
$\Delta\Omega=\Omega_{\rm eq}-\Omega_{\rm pole}=\alpha\Omega_{\rm eq}$.

Measuring DR is challenging, because it acts as a subtle perturbation of the mean rotation
rate. For stars other than the Sun different techniques have been used.
\citet{Reiners2002a} applied the Fourier transform method to Doppler broadened spectral 
lines. Doppler Imaging \citep{Donati1997} traces active regions located at different 
latitudes to infer the absolute surface shear $\Delta\Omega$. Furthermore, DR can also be
measured from asteroseismology \citep{Gizon2004,Lund2014}. Classical spot models also
supply a measure on DR, with recent studies being, e.g., \citet{Lanza2014,Bonanno2014}.
Using the short-term Fourier-transform \citet{Vida2014} provided measurements of activity
cycles of fast rotating Kepler stars attributing the observed change in rotation period
to DR. Another approach was presented by \citet{Reinhold2013b} by measuring different
periods in high-precision light curves. This method only yields absolute values of the
shear but provides no information of the sign, because the latitudes of the active regions
are usually unknown.

\begin{figure}
  \resizebox{\hsize}{!}{\includegraphics{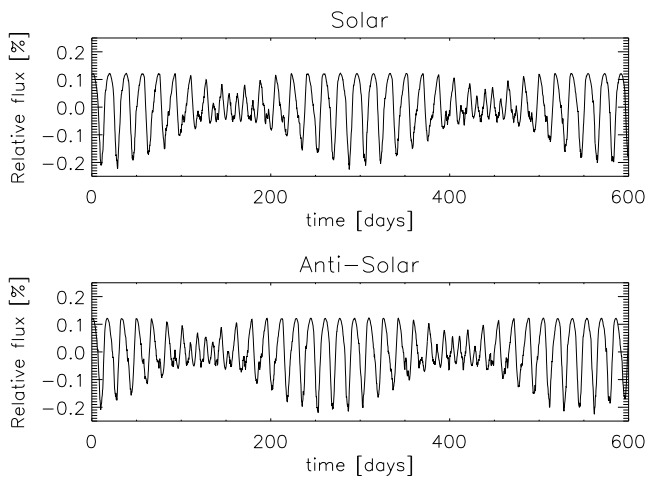}}
  \caption{\textit{Upper panel:} Simulated light curve of a star with two spots
  and an equator-to-pole shear $\alpha=0.085$ (solar-like DR) according to
  eq.~\ref{drotlaw}. \textit{Lower panel:} Same parameters as above but with a negative
  value of $\alpha=-0.085$ instead (antisolar DR).}
  \label{example_lcs}
\end{figure}

Fig.~\ref{example_lcs} illustrates that it is difficult to tell from the light curve
whether the star exhibits solar or antisolar DR. Both light curves were drawn from the
same parameters and show an almost identical beating pattern, with the exception that
$\alpha$ is positive (negative) in the upper (lower) panel, respectively. Thus, visual
inspection of the light curve does not immediately reveal the underlying rotational
behavior.

Recently, 3D numerical simulations have tabled the question which physical conditions
lead to solar-like or antisolar differential rotation, and which parameters determine the
transition region \citep{Guerrero2013, Gastine2014, Kaepylae2014, Karak2014}. A
theoretical explanation for the existence of antisolar DR is provided by
\citet{Kitchatinov2004} favoring the explanation of fast meridional motion. Nevertheless,
observations of antisolar DR are sparse and have only been claimed in some K giants so
far. \citet{Strassmeier2003} found strong signatures of antisolar DR on the K2-giant HD
31993 supplying the value $\alpha=-0.125\pm0.05$. \citet{Weber2005} found marginal
evidence for antisolar DR in several K giants. \citet{Kovari2007} detected antisolar DR
of the K1 giant $\sigma$ Geminorum providing a relative shear of $\alpha \approx
-0.022\pm0.006$. Later studies of the same object approved antisolar DR, almost
twice as strong as the earlier value \citep{Kovari2014}. \citet{Vida2007} observed
antisolar DR in the K0 giant UZ libr\ae{} using cross-correlation of consecutive Doppler
images obtained in the years 1998 and 2000, yielding $\alpha \approx -0.03$ and $\alpha
\approx -0.004$, respectively. \citet{Dikpati2011} analyzed the stability of toroidal
magnetic fields and differential rotation and found that kG-fields (i.e. activity) are
only compatible with small shear if the DR is antisolar. The finding agrees with the small
values observed. Furthermore, \citet{Ammler2012} applied the Fourier transform method to a
sample of more than 110 A--F field stars. For the majority of stars in their sample these
authors either found no signature of DR or very small values consistent with rigid
rotation. For 10 stars significant DR $(0.10 < \alpha < 0.50)$ was detected. Five stars in
their sample showed signatures which can be explained by antisolar DR. Nevertheless, the
authors interpreted the measurements as rigid rotators with cool polar caps.

The goal of this article is to present a new method to measure the sign of DR from
high-precision light curves. We compute their Lomb-Scargle periodograms and show how
different periods can be associated with higher or lower latitudes using overtone
amplitudes. No explicit spot modeling is needed! We explain the method in detail, present
results for simulated light curves, and discuss the problems when applying this technique
to Kepler data. The paper is organized as follows: In Sect.~\ref{simulations} we motivate
the parameter space of the simulated light curves. Section \ref{method} contains a
detailed description of the method. The application of the method to simulated and real
data is presented in Sect.~\ref{results}. The last section is dedicated to a detailed
discussion of the results.

\section{Simulations}\label{simulations}
In order to test our method we simulated different sets of differentially rotating stars.
Each set consists of 100 light curves and was generated by our code MODSTAR
\citep{Reinhold2013a}. The code creates a spherical surface of a rotating spotted star.
Each surface element has a fixed intensity value. Spots are modeled as circular areas with
a spot-to-photosphere contrast of 0.67, which roughly corresponds to the solar
penumbra-to-photosphere contrast. For simplicity, umbra and penumbra structures of
star spots as well as non-circular spot shapes were neglected.

The crucial parameters to describe latitudinal shear are the equatorial rotation period
$P_{\rm eq}=2\pi/\Omega_{\rm eq}$, the equator-to-pole shear $\alpha$ , and the spot
latitudes (see eq.~\ref{drotlaw}). Further parameters of interest are the number of spots
and the stellar inclination angle $i$. All light curves were created using the following
parameters, which were chosen at random for each light curve. The stellar inclination 
ranges from $30\degree < i < 80\degree$, the equatorial rotation period from 
$1 < P_{\rm eq} < 20$ days, and the equator-to-pole shear covers $0.05 < \alpha < 0.30$.
We emphasize that all simulated stars exhibit a positive $\alpha$, and thus solar-like DR!

The first model set consists of 100 stars with only two spots at fixed latitudes
$\theta_1=10\degree$ and $\theta_2=60\degree$. The second set contains 100 model stars
with 10 spots located at random latitudes between $-30\degree < \theta < 90\degree$. The
number of spots and their latitudes is the only difference between the two sets. The spots
can cover the full longitude range $-180\degree < \varphi < 180\degree$, with radii
between $4\degree < r < 6\degree$. For the 10-spot models these radii correspond
to a spot coverage of $\sim1-3$\%. Additionally, spots are allowed to overlap without
further contrast reduction. Each light curve covers a total length of 1500 days, with a
sampling rate of 4 points/day.

The spots in our simple models have infinite lifetimes, i.e., their radii do not change
with time. Spot evolution can also be implemented in the models by increasing or
decreasing the spot radii. In Sect.~\ref{results} we test the robustness of our method
under various perturbations, especially probing the effect of spot evolution. Since there
are many ways of realizing spot evolution, we chose a simple approach by changing the spot
radii when creating the light curves. At the beginning of the simulation the program
chooses at random whether all spots should be increased or decreased. Over a course of 30
days, all spot radii are then successively changed by 0.4$\degree$ in total. Every 30
days, the program chooses again whether the spots should be increased or decreased. Owing
to the method it happens that spot radii drop below $0\degree$. In such cases the spots
are forced to increase again during the next 30 days period. Both the 30-day interval as
well as the amount of 0.4$\degree$ growth or decay in radii was chosen such that the light
curves visually exhibit similar shape to what can be found in Kepler data.

The first model set only contains stars with two spots on their surface. These sets
might not be very realistic, but are sufficient to probe the signature of higher and
lower latitude spots in the light curve. This set is ideal to present the power of our
method, and to focus on the primary goal of detecting the correct sign of $\alpha$. The
other parameters were chosen to be comparable with Kepler data. We chose an upper limit of
20 days or the equatorial rotation period, because many Kepler stars exhibit very stable
rotation periods in this time regime. Longer periods usually become unstable due to spot
evolution, stronger differential rotation, or instrumental effects. From previous
simulations we know that DR is hard to detect in stars with small equator-to-pole shear
\citep{Reinhold2013a}, so a lower limit of $\alpha > 0.05$ was imposed. The inclination is
limited to $i > 30\degree$, because our method tends to fail for lower values.
Furthermore, the total length of the light curves in comparable to Kepler's observing time
span.

\section{Method}\label{method}
We compute the generalized Lomb-Scargle periodogram \citep{Zechmeister2009}
of the full light curves, identify the 30 highest peaks, and sort them by their peak
heights\footnote{We use the normalization from eq.~22 in \citet{Zechmeister2009}.} in
descending order. The highest-peak period is termed $P_1$ and represents the most
significant period in the data. To measure the amount of latitudinal DR we search for 
periods adjacent to $P_1$ in the range of $P_1\pm30\,\%$ with peak heights greater than
50. This value represents a lower limit of the significance level of the period in the
data, and was chosen after by-eye inspection of several periodograms. The second highest
period satisfying the above conditions is termed $P_2$, the third highest peak $P_3$, and
so on. 

Ideally these periods belong to certain spot rotation periods, i.e., to spots rotating at
certain latitudes. For the 2-spot model we found that this assignment is true for almost
all cases (see Sect.~\ref{2spots}). Unfortunately, for the 10-spot models 
(Sect.~\ref{10spots}) this association is not valid anymore, because not every spot is
resolved by an individual peak. Furthermore, in real data the number of spots on the
surface is unknown, and one has to assign the peaks to spots (or spot groups) on the
surface to get a measure of the latitudinal DR. Naturally one would use the highest-peak
periods $P_1$ and $P_2$ to guarantee that these are really present in the data. But no
matter which peaks are used to get a measure on DR, they only provide an \textit{absolute}
value of the surface shear, because there is no information about the associated spot
latitudes. In the following we show that latitudinal information can be achieved by taking
into account the first harmonics (half periods) of the selected periods.

The light curve signature of a single star spot rotating in and out of view depends on its
radius, the intensity contrast with its surrounding, the stellar inclination angle, and
the spot latitude. In our model we assume circular spot shape and neglect any
temperature gradient within the spot. Our method is based on the outcome of the
simulations that -- for a wide parameter range  -- lower latitude spots generate less
sine-shaped light curves compared to higher latitude spots. As a consequence, additional
harmonics of the primary period are required for a proper fit. Since the goodness of a
sine fit is equivalent to the peak height in the periodogram, one can achieve information
about the spot latitude by comparing the heights of selected peaks and their harmonics. We
search the periodogram for the first harmonic of \textit{all} periods $P_k,\ k=1,2,...$
satisfying the above conditions, because in some cases the harmonics of $P_1$ and $P_2$
cannot be detected. If present, the harmonics are termed $P_k'$ and should lie within
1\,\% of the half periods $P_k/2$, i.e., they should satisfy $0.99 P_k/2 < P_k' < 1.01
P_k/2$. The corresponding peak heights of the periods $P_k$ are termed $h(P_k)$ and of the
harmonics $h(P_k')$, respectively. For all pairs ($P_k,P_k'$) satisfying the above
relations we compute the peak-height-ratios (PHR) $r_k:=h(P_k')/h(P_k)$. Comparing
different ratios now reveals information about the spot latitude. The relation $r_k >
r_{k+1}$ means that the relative power of the harmonic compared to the power of the
corresponding period is higher for $k$ than for $k+1$. In this case, we associate
the period $P_k$ with a lower latitude spot compared to the spot latitude associated with
$P_{k+1}$. We define
\begin{equation}\label{eq2}
  \left.
  \begin{array}{c}
    \hspace{2cm}
    P_{\rm low} := P_k \\
    \hspace{2cm}
    P_{\rm high} := P_{k+1}
  \end{array}
  \right\} \qquad \text{if} \qquad r_k > r_{k+1}
\end{equation}
and 
\begin{equation}\label{eq3}
  \left.
  \begin{array}{c}
    \hspace{2cm}
    P_{\rm low} := P_{k+1} \\
    \hspace{2cm}
    P_{\rm high} := P_k
  \end{array}
  \right\} \qquad \text{if} \qquad r_{k+1} > r_k.
\end{equation}
Since we sorted the periods according to their peak heights, we start to compare the PHR
with the lowest $k$-value, $k_{\rm min}$, satisfying all conditions, to infer the rotation
law. In most cases, $k_{\rm min}=1$ is used if $P_1'$ is detectable. If $P_{\rm low} <
P_{\rm high}$ for $k=k_{\rm min}$ the star exhibits solar-like DR. If the opposite
relation is true, antisolar DR is present. Thus, comparing the PHR and the corresponding
periods yields the sign of DR!

In some cases contradictory results are returned by comparing the PHR for different
$k$-values, e.g., when the comparison for $k=1$ reveals solar-like DR, and for $k=2$
antisolar DR, or vice versa. This happens because the routine matches all peaks in the
range of the rotation period to their harmonics, and associates them with spot rotation
rates at different latitudes. Because of the fact that not each and every single peak can
be associated with a certain spot rotation rate, in some cases the periods are associated
with the wrong latitudes. To minimize this effect, we only use the relation for 
$k=k_{\rm min}$, because $k_{\rm min}$ uses the two highest peaks with detectable
harmonics, for comparison. 

For the 10-spot models we checked under which conditions a negative sign of DR is
returned. To minimize the number of wrong detections we imposed some constraints on the
program. For stars with too many individually resolved periods $P_k$, we restricted the
search to $k \leq 5$. Furthermore, the ratios $r_k$ and $r_{k+1}$ must differ by at least
10\,\%. The largest contribution of wrong detections results from choosing peaks very
close to each other, so we forced the program to only match peaks which are separated by
at least five points on the frequency grid. Relaxing this condition has a large impact on
the number of detection in real data (see Sect.~\ref{kepler}), and is discussed in Sect.
\ref{discussion}.
\begin{figure}
  \resizebox{\hsize}{!}{\includegraphics{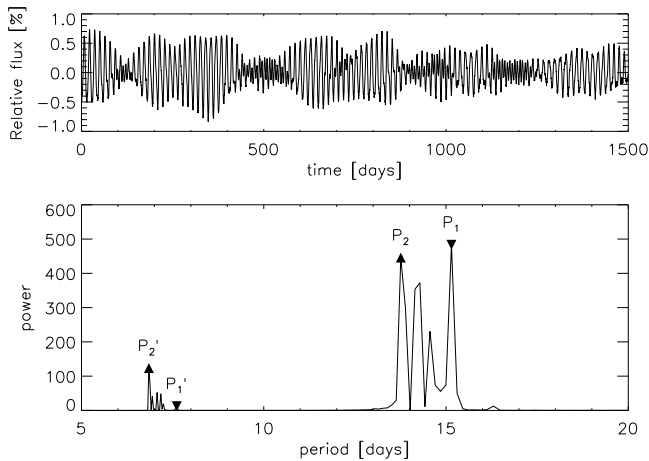}}
  \caption{\textit{Upper panel:} Light curve of a randomly selected star from the 10-spot
  models without spot evolution. \textit{Lower panel:} Periodogram of the above light
  curve. The peaks of the pairs ($P_1,P_1'$) and ($P_2,P_2'$) are clearly visible
  and emphasized by black upside down and normal triangles, respectively.}
  \label{method_lc38_10spots}
\end{figure}
To illustrate the method Fig.~\ref{method_lc38_10spots} shows the periodogram of a
randomly selected light curve from the 10-spot model without spot evolution. In this model
the periods and harmonics for $k=1,...,4$ satisfy all conditions and are collected in
Table~\ref{fig2_table}. To infer the rotation law the peaks of the pairs ($P_1,P_1'$) and
($P_2,P_2'$) are selected and emphasized by black upside down and normal triangles,
respectively. We detected the periods $P_1=15.15$\,d, $P_2=13.76$\,d, $P_1'=7.61$\,d,
$P_2'=6.85$\,d, and the ratios $r_1=0.026$ and $r_2=0.277$. Since $r_2 > r_1$ the period
$P_2$ belongs to a lower latitude spot (see eq.~\ref{eq3}), and the relation $P_2 < P_1$
reveals solar-like DR!
The original spot rotation periods $P_{\rm spot}$ of this model lie in the range 
13.71\,d $< P_{\rm spot} < 16.47$\,d, showing that the periodogram detects different
periods in the right range. Further parameters of the picked model are $i=60.6\degree$ and
$\alpha=0.168$.
\begin{table}
  \begin{center}
    \begin{tabular}{ccccc}
      $k$ & 1 & 2 & 3 & 4 \\
      \hline
      $P_k$ & 15.15 & 13.76 & 14.28 & 14.56 \\
      $P_k'$ & 7.61 & 6.85 & 7.07 & 7.25 \\
      $r_k$ & 0.026 & 0.277 & 0.141 & 0.085 \\
    \end{tabular}
  \end{center}
  \caption{Periods $P_k$, harmonics $P_k'$, and ratios $r_k$ for $k=1,...,4$ satisfying
  all conditions for the model star in Fig.~\ref{method_lc38_10spots}.}
  \label{fig2_table}
\end{table}

\section{Results}\label{results}
In this section we present the results of the application of our method to the different
artificial and real data sets. To quantify the results we provide the distribution of the
observed relative differential rotation $\alpha_{\rm obs}$, which we define as
\begin{equation}
  \alpha_{\rm obs}:=\frac{P_{\rm high}-P_{\rm low}}{P_{\rm high}},
\end{equation}
with the periods $P_{\rm high}$ and $P_{\rm low}$ from eqn. (\ref{eq2}) and (\ref{eq3}).
This quantity automatically provides the \textit{false-positive rate} of our method, being
the number of stars with negative $\alpha_{\rm obs}$ values. Negative values of
$\alpha_{\rm obs}$ result from associating certain periods with the wrong latitudes as
discussed in Sect.~\ref{method}. Again, we remind the reader that all simulated stars
exhibit $\alpha>0$. 

As we will show in the following, the false-positive rate is rather small for
\textit{unperturbed} data. The above sets of light curves are noise-free, the spots do not
evolve with time, and -- compared to real data -- come without gaps or instrumental
effects. To test the robustness of our method we include different perturbations in the
light curves. We found that spot evolution has the largest effect, so for each set we
varied the spot radii with time (see Sect.~\ref{simulations}) using the same parameters as
in the unperturbed case, and considered different levels of Poisson noise.

To quantify the impact of each effect we compare the distribution of $\alpha_{\rm obs}$
to the distribution of the measurable shear of the model. Therefore, we define
$\alpha_{\rm model}:=(P_{\rm max}-P_{\rm min})/P_{\rm max}$, with 
$P_{\rm max}=\max(P(\theta))$ and $P_{\rm min}=\min(P(\theta))$, being the maximum and
minimum periods of the model, respectively. By definition it holds 
$\alpha_{\rm obs}\leq\alpha_{\rm model}$, because the periods $P_{\rm high}$ and 
$P_{\rm low}$ do not necessarily belong to the highest and lowest latitude spots,
respectively. For each model set we show how well the original spot periods are
recovered, directly providing the false-positive rate of our method.

\subsection{2-spot models}\label{2spots}
\begin{figure*}
  \centering
  \includegraphics[width=17cm]{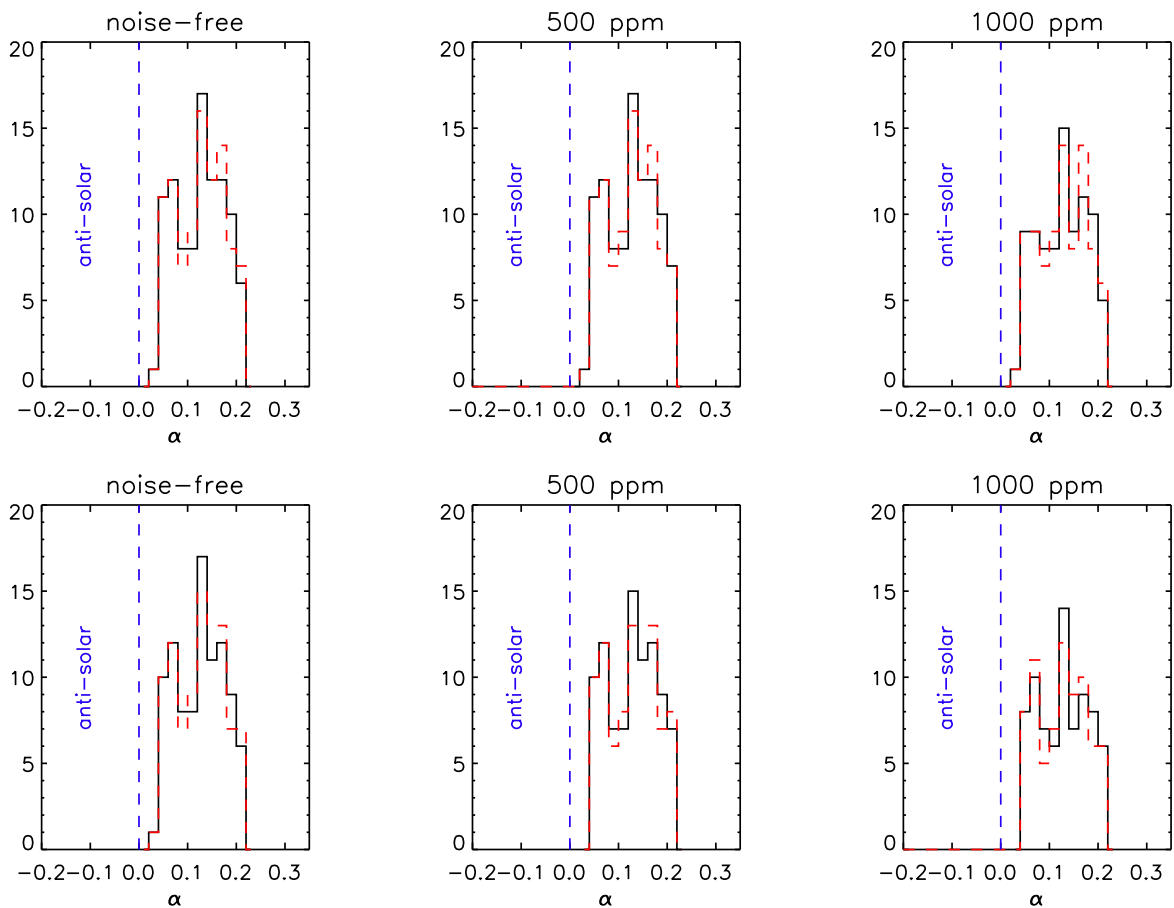}
  \caption{Distributions of $\alpha_{\rm model}$ (solid black) and $\alpha_{\rm obs}$
  (dashed red) of the 2-spot model for different noise levels. \textit{Upper panel:} No
  spot evolution, \textit{Lower panel:} Spot evolution included. The dashed blue line
  indicates $\alpha_{\rm obs}=0$.}
  \label{alpha_out_2spots}
\end{figure*}

The distributions of $\alpha_{\rm model}$ and $\alpha_{\rm obs}$ are shown in Fig.
\ref{alpha_out_2spots} in solid black and dashed red, respectively. The upper panel shows
the distributions without spot evolution, whereas in the lower panel spot evolution is
included in the models. In both panels, the noise level increases from left to right.

For non-evolving spots we find $\alpha_{\rm obs} > 0$ in 97 of 100 cases in the
noise-free models. For the 3 remaining stars the conditions were not fulfilled, i.e., the
false-positive rate is zero! Almost the same result was found when adding 500 ppm Poisson
noise to the light curves, with 97 solar-like and 1 antisolar detections. Adding 1000 ppm
white noise yields solar-like DR for 85 stars, and no detection for the remaining 15
light curves. Adding even higher noise seems unreasonable, because the periodicity in the
light curves visible to the naked eye gets corrupted. Although the main rotation
periods are still detectable, their harmonics cannot be identified correctly, because the
large noise is interpreted as short periods in the periodograms. Such noisy light curves
have not been considered when searching for DR in Kepler data \citep{Reinhold2013b}.

We also considered the same models including spot evolution. For the noise-free, the
500 ppm, and the 1000 ppm models we find 94, 90, and 74 solar-like detections,
respectively. Only one case of antisolar rotation was detected in the 1000 ppm model. In
all cases we find good agreement between the model periods and the recovered values.
Furthermore, the false-positive rate is zero for almost all cases supporting the power of
the method.

\subsection{10-spot models}\label{10spots}
\begin{figure*}
  \centering
  \includegraphics[width=17cm]{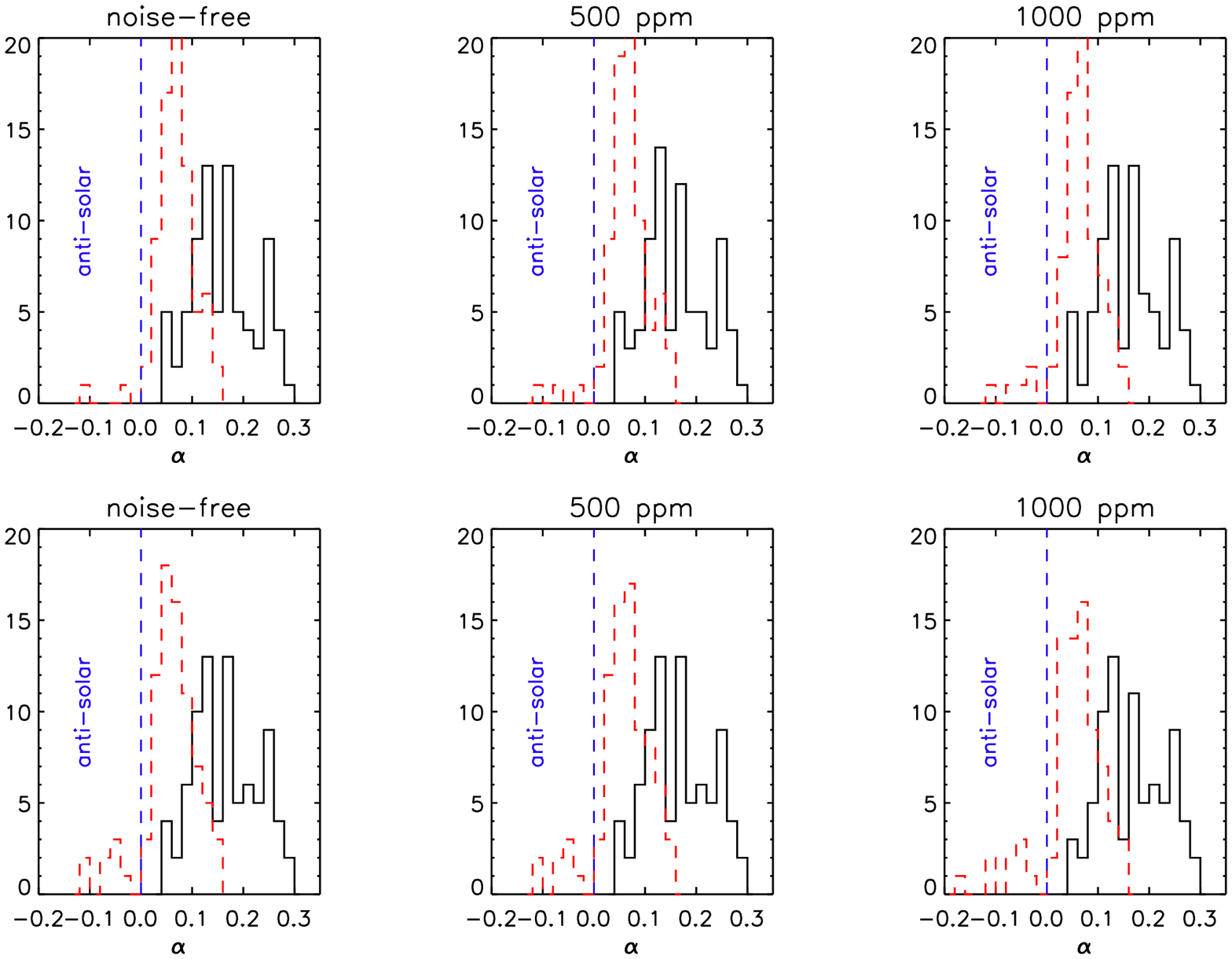}
  \caption{Distributions of $\alpha_{\rm model}$ (solid black) and $\alpha_{\rm obs}$
  (dashed red) of the 10-spot model for different noise levels. \textit{Upper panel:} No
  spot evolution, \textit{Lower panel:} Spot evolution included. False-positive detections
  exhibiting $\alpha_{\rm obs}<0$ are located left of the dashed blue line (indicating 
  $\alpha_{\rm obs}=0$).}
  \label{alpha_out_10spots}
\end{figure*}

In Fig.~\ref{alpha_out_10spots} we show the distributions of $\alpha_{\rm model}$ and
$\alpha_{\rm obs}$ of the 10-spot models in solid black and dashed red, respectively.
Again, distributions in the upper panel are without spot evolution, whereas spot evolution
is included in the lower panel, and the noise level increases from left to right.

The distributions of $\alpha_{\rm obs}$ look very different to the ones of the 2-spot
models, with $\alpha_{\rm obs}$ being much lower than $\alpha_{\rm model}$, on average.
The shear is underestimated in many cases, because the stars exhibit more than only two
significant periods. The match between $\alpha_{\rm obs}$ and $\alpha_{\rm model}$ can
be improved by picking other periods, but our primary goal was to detect the correct sign
of $\alpha$, and not to recover the total amount of shear as good as possible. The values
of $\alpha_{\rm obs}$ should rather be considered as lower limits of the in principle
measurable shear. 

The false-positive rates for the different cases are as follows. For the models without
spot evolution we find 2 of 78, 3 of 78, and 5 of 77 antisolar cases for the noise-free,
the 500 ppm, and the 1000 ppm models, respectively. Including spot evolution the number of
wrong associations increases to 8 of 83, 8 of 82, and 9 of 78 for the noise-free, the 500
ppm, and the 1000 ppm models, respectively. This result clearly shows that spot evolution
has a much bigger effect on the false-positive rate than adding Poisson noise to the
light curves.

According to the above results our method exhibits a false-positive rate of less
than $\sim12$\%.
Relaxing the constraints imposed on the method, especially allowing for smaller peak
separations (see Sect.~\ref{method}), leads to higher false-positive rates, as discussed
in Sects.~\ref{kepler} and \ref{discussion}.

\subsection{Kepler data}\label{kepler}
In this section we apply our method to a sample of 50 selected Kepler stars. As
shown previously our method performs well for simulated data. We were curious to see
whether it is possible to detect the sign of DR in real data in general, especially
answering the question for how many stars antisolar DR is revealed by our method.
Therefore, we selected a hand-picked sample of \textit{solar-like} stars from the
Kepler archive. Although solar-like DR is expected for these stars, the underlying
differential rotation law is still unknown.

The selected stars should hold the following criteria: surface gravities $\log g > 3.5$,
color index $0.60 < B-V < 0.92$ (roughly corresponding to spectral type G0--G9 using the
$g-r$ to $B-V$ color transform from \citet{Jester2005}), and Kepler magnitude $kepmag <
13$, leaving roughly 5,700 stars.
Since we are only interested in rotation, the above sample was further reduced by
discarding eclipsing binaries\footnote{http://keplerebs.villanova.edu/} and 
KOIs\footnote{http://archive.stsci.edu/kepler/koi/search.php}. Two stars in the sample
were characterized as hybrid pulsators \citep{Uytterhoeven2011}, and one star as $\gamma$ 
Doradus pulsator \citep{Tkachenko2013}. These stars were also discarded.

To minimize the effect of possible instrumental flaws we only selected stars with strong
intrinsic variability, expressed in terms of the \textit{variability range} 
$R_{\rm var} > 0.3\,\%$ \citep{Basri2011}. Additionally, the most significant period in
the periodogram should hold $1 < P_1 < 25$ days. These limits should guarantee a stable
rotational modulation over many rotation cycles.

We only used Kepler Q1--Q14 data, which were consistently reduced with the PDC-MAP
pipeline (version 2.1--3.0). Q15--Q17 data have not been considered, because these data
were reduced with the msMAP pipeline, which applies a 20 day high-pass filter to certain
quarters, thus affecting the signal of slower rotators. The light curves were stitched
together in a simple way by dividing each quarter by its median and subtracting unity. The
light curves were smoothed with a 30-day window to remove remaining long-term trends,
likely caused by the instrument.

Furthermore, we visually inspected all light curves and their periodograms. We only
selected stars with multiple periodogram peaks, at least one visible harmonic, and clearly
visible rotational modulation in the light curve. The above criteria and the visual
inspection left only 50 stars in total.

Our method described in Sect.~\ref{method} was applied using the same constraints
as for the simulated data. We found that the separation of the two peaks $P_1$ and $P_2$
on the frequency grid has a strong influence on the determination of the correct sign of
DR. Therefore, we define 
\begin{equation}
  \Delta f := |1/P_1 - 1/P_2|
\end{equation}
To have a dimensionless quantity for the peak separation, we divide this frequency
difference by the minimum frequency $f_{\rm min}=1/\max(\rm time)$ of the periodogram. In
the following, the number of points on the frequency grid separating two peaks is called
the peak separation.
Forcing the program only to pick peaks with $\Delta f/f_{\rm min} \geq n$ with $n\geq 2$
rules out certain peak combinations. This happens, e.g., when more than two distinct
rotation peaks are found, but only the two closest peaks exhibit a detectable harmonic.
For this reason the total number of detections drops when increasing the minimum peak
separation (s. second column in Table~\ref{peaksep_table}).

\begin{figure*}
  \centering
  \includegraphics[width=17cm]{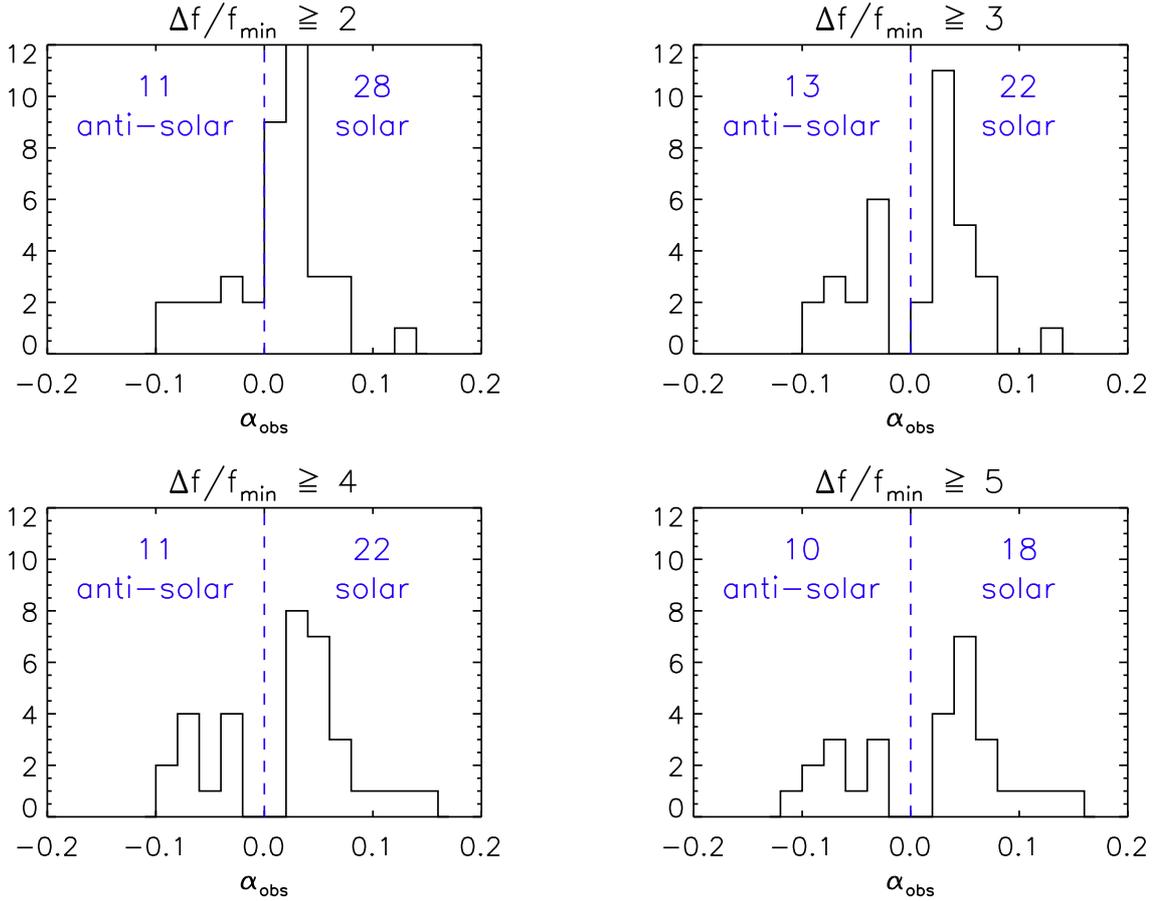}
  \caption{Distribution of $\alpha_{\rm obs}$ of the 50 Kepler stars forcing different
  minimum peak separations. The dashed blue line indicates rigid rotation ($\alpha_{\rm
  meas}=0$).}
  \label{alpha_kepler}
\end{figure*}

Table~\ref{peaksep_table} summarizes the results and Fig.~\ref{alpha_kepler} shows the
distribution of $\alpha_{\rm obs}$ using the different constraints. Increasing the
minimum peak separation significantly reduces the total number of detections, especially
the cases where the highest peaks have been chosen ($k_{\rm min}=1$). The number of
detections of antisolar-type differential rotation varies from 28--37\,\%.

To put the Kepler results into context we applied our method to the 10-spot models
including spot evolution and 1000 ppm white noise using the same peak separations as
above. We found that reducing the minimum peak separation from five to two frequency bins
increases the percentage of false-positives from 11.3\,\% to 20\,\%. The number of cases
with $k_{\rm min}=1$ fluctuates between 78--91\,\%. This might suggest that the result for
$\Delta f/f_{\rm min} \geq 2$ is least reliable, although $k_{\rm min}=1$ was used in all
cases. Such split peaks with small peak separation are likely caused by spot evolution.

\begin{table}
  \begin{center}
  \begin{tabular}{ccccc}
    Peak sep. & \# detections & \# detections & Solar & Antisolar \\
    $\Delta f/f_{\rm min} $ & (total) & ($k_{\rm min}=1$) & $\alpha >0$ & $\alpha <0$ \\
    \hline
    $\geq2$ & 39 & 39 & 28 & 11 \\ 
    $\geq3$ & 35 & 28 & 22 & 13 \\ 
    $\geq4$ & 33 & 20 & 22 & 11 \\ 
    $\geq5$ & 28 & 13 & 18 & 10 \\ 
  \end{tabular}
  \end{center}
  \caption{Number of solar and antisolar detections out of the 50 Kepler stars. $\Delta
  f/f_{\rm min}$ denotes the minimum peak separation. The second column shows the total
  number of detections. Thereof, the number of detections with $k_{\rm min}=1$ is given
  in the third column. The last two columns denote the number of solar and antisolar
  detections, respectively.}
  \label{peaksep_table}
\end{table}

\section{Discussion}\label{discussion}
As shown in the previous section, our method performs well for simulated data, especially
the results of the 2-spot models are striking. The detection rate for the 10-spot models
is still high, although the periodograms exhibit more than only two significant rotation
peaks, as do their harmonics (see Fig.~\ref{method_lc38_10spots}).

The false-positive rate of our method depends on the noise level, the minimum peak
separation, and the presence of spot evolution. We found that our method is very robust
against adding pure white noise to the light curves, in agreement with other studies
\citep{Reinhold2013a}. Adding higher white noise to the light curves decreases the total
number of detections, but does not significantly increase the false-positive rate.
Spot evolution was found to be the major contributor of spurious detections. The largest
rate was found for the 10-spot models including spot evolution and having small peak
separations, increasing up to 20\,\% for the highest noise levels. In certain cases very
short-lived spots can produce split rotation peaks, and thus mimic differential rotation.

In real data this effect becomes more important, because spot lifetimes are difficult to
assess. In some situations spot evolution can barely be distinguished from DR, making the
determination of the sign of DR is even harder. In general, periodograms of Kepler stars
exhibit many more peaks, with some of them being spurious detections. Furthermore, the
periodogram strongly depends on the length of time series. Since the peak width is
proportional to the inverse of the total observing time span, a certain length of
high-quality photometry is required to resolve individual peaks. But also the analysis of
two different light curve segments of equal length may change the result. Spot evolution
and differential rotation may be more ``visible'' in one segment than in the other, as
revealed by their periodograms, either showing a single or multiple clear peaks,
respectively. This effect immediately connects to the different peak separations (see 
Fig.~\ref{alpha_kepler}). Computing the periodogram of the full time series averages out
these effects. Therefore, analyzing different light curve segments might increase the
sample of suitable candidates to test this method.

Unfortunately, the false-positive rate of our method cannot be transferred directly to
Kepler data, because the percentage of real antisolar rotators is unknown. It should
rather be interpreted as the number of cases where true solar rotators are misclassified
as antisolar rotators, and vice versa. Since the light curves and periodograms of solar
and antisolar rotators exhibit similar shapes, one can argue that both cases should be
treated equally by the method, i.e., the rate of stars with wrongly determined sign of
DR should be the same on either side. The measured number of antisolar detections then
consists of the fraction of real antisolar rotators, which are actually detected as such,
plus the number of migrated solar-like stars, misclassified as antisolar rotators. Thus,
we define
\begin{eqnarray}
  \hspace{2cm}
  S' &=& (1-f)S+fA \\
  \hspace{2cm}
  A' &=& (1-f)A+fS.
\end{eqnarray}
$S'$ and $A'$ are the measured solar and antisolar detections, respectively, $S$ and $A$
the number of real solar and antisolar rotators, respectively, and $f=0.2$ the maximum
false-positive rate. Solving this system of linear equations yields
\begin{eqnarray}
  \hspace{2cm}
  A &=& \frac{(1-f)A'-fS'}{1-2f} \\
  \hspace{2cm}
  S &=& A'+S'-A.
\end{eqnarray}
Applying this solution to the numbers given in Table~\ref{peaksep_table} we detect solar
DR in 21--34 and antisolar DR in 5--10 stars of our sample of 50 stars. 
Table~\ref{DR_table} lists the KIC numbers of our sample of 50 Kepler stars, providing
their differential rotation law found by our method for the different peak separations.
Most of the stars are consistently found as solar (s) or antisolar (a) rotators, whereas
some switch from one to the other. Additionally, one can see that forcing a larger peak
separation leads to a non-detection (-) for some stars.
\begin{table}
  \begin{center}
    \begin{tabular}{|c|c|c|c|c|}
\hline
\multirow{2}{*}{KIC} & \multicolumn{4}{|c|}{$\Delta f/f_{\rm min} \geq$} \\
\cline{2-5}
 & 2 & 3 & 4 & 5 \\
\hline
1724975 & s & s & s & s \\
\hline
3352189 & a & a & - & - \\
\hline
3837480 & s & s & s & - \\
\hline
3862793 & - & - & - & - \\
\hline
4557559 & s & s & s & a \\
\hline
4639291 & a & a & a & a \\
\hline
4677300 & s & s & s & s \\
\hline
4758719 & a & a & a & a \\
\hline
4829624 & - & - & - & - \\
\hline
5032950 & s & s & s & s \\
\hline
5091941 & s & s & s & s \\
\hline
5165799 & s & s & s & s \\
\hline
5621105 & s & s & s & s \\
\hline
6038355 & a & a & a & a \\
\hline
6064684 & s & s & s & s \\
\hline
6145471 & s & s & s & - \\
\hline
6182131 & a & a & a & a \\
\hline
6633742 & - & - & - & - \\
\hline
6976397 & s & a & a & - \\
\hline
7201168 & s & s & s & s \\
\hline
7385442 & a & a & a & a \\
\hline
7434544 & s & s & s & s \\
\hline
7601913 & a & - & - & - \\
\hline
7678060 & s & - & - & - \\
\hline
7739728 & s & - & - & - \\
\hline
7950229 & - & - & - & - \\
\hline
8160546 & s & s & s & s \\
\hline
8180580 & - & - & - & - \\
\hline
8226464 & a & a & a & a \\
\hline
8299279 & s & s & s & s \\
\hline
8311074 & s & a & a & a \\
\hline
8457754 & s & s & s & a \\
\hline
8883245 & - & - & - & - \\
\hline
8903403 & s & a & a & a \\
\hline
8957218 & s & s & s & s \\
\hline
9002888 & s & s & s & s \\
\hline
9161418 & - & - & - & - \\
\hline
9172729 & s & s & s & s \\
\hline
9222547 & - & - & - & - \\
\hline
9272360 & - & - & - & - \\
\hline
9389733 & a & a & a & - \\
\hline
9693116 & - & - & - & - \\
\hline
9715600 & s & s & s & s \\
\hline
10544691 & s & s & s & s \\
\hline
10797565 & a & a & a & - \\
\hline
11619829 & a & a & s & s \\
\hline
11909695 & - & - & - & - \\
\hline
12217823 & s & s & - & - \\
\hline
12554089 & s & - & - & - \\
\hline
12691333 & s & s & s & s \\
\hline
\end{tabular}

  \end{center}
  \caption{Solar (s) and antisolar (a) detections for the selected 50 Kepler stars
  depending on the peak separation.}
  \label{DR_table}
\end{table}

As discussed above this new method performs well in our simulations, but should be used
with caution in real data. Although we tried to successively approach the shape of real
light curves by increasing the number of spots, probing the effect of spot evolution, and
considering different noise scenarios, some effects have not been considered. The spot
evolution as implemented in our model may be over-estimating real spot lifetimes.
Furthermore, instrumental artifacts in the quarterly delivered Kepler data, and their
reduction by the PDC-MAP pipeline certainly affect the overall light curve shape, and thus
the period determination. 

This number of antisolar detections seems quite large, but reflects the outcome of the
method. Since the light curves created from the 10-spot models look very similar to real
Kepler data, we are optimistic that our findings of antisolar DR are real, although
independent measurements to either prove or disprove our results are urgently needed.

\acknowledgements
TR acknowledges the support from Prof. Laurent Gizon.

\bibliography{biblothek}

\begin{thebibliography}{26}
\expandafter\ifx\csname natexlab\endcsname\relax\def\natexlab#1{#1}\fi

\bibitem[{{Ammler-von Eiff} \& {Reiners}(2012)}]{Ammler2012}
{Ammler-von Eiff}, M. \& {Reiners}, A. 2012, \aap, 542, A116

\bibitem[{{Basri} {et~al.}(2011){Basri}, {Walkowicz}, {Batalha}, {Gilliland},
  {Jenkins}, {Borucki}, {Koch}, {Caldwell}, {Dupree}, {Latham}, {Marcy},
  {Meibom}, \& {Brown}}]{Basri2011}
{Basri}, G., {Walkowicz}, L.~M., {Batalha}, N., {et~al.} 2011, \aj, 141, 20

\bibitem[{{Bonanno} {et~al.}(2014){Bonanno}, {Fr{\"o}hlich}, {Karoff}, {Lund},
  {Corsaro}, \& {Frasca}}]{Bonanno2014}
{Bonanno}, A., {Fr{\"o}hlich}, H.-E., {Karoff}, C., {et~al.} 2014, \aap, 569,
  A113

\bibitem[{{Dikpati} \& {Cally}(2011)}]{Dikpati2011}
{Dikpati}, M. \& {Cally}, P.~S. 2011, \apj, 739, 4

\bibitem[{{Donati} \& {Collier Cameron}(1997)}]{Donati1997}
{Donati}, J.-F. \& {Collier Cameron}, A. 1997, \mnras, 291, 1

\bibitem[{{Gastine} {et~al.}(2014){Gastine}, {Yadav}, {Morin}, {Reiners}, \&
  {Wicht}}]{Gastine2014}
{Gastine}, T., {Yadav}, R.~K., {Morin}, J., {Reiners}, A., \& {Wicht}, J. 2014,
  \mnras, 438, L76

\bibitem[{{Gizon} \& {Solanki}(2004)}]{Gizon2004}
{Gizon}, L. \& {Solanki}, S.~K. 2004, \solphys, 220, 169

\bibitem[{{Guerrero} {et~al.}(2013){Guerrero}, {Smolarkiewicz}, {Kosovichev},
  \& {Mansour}}]{Guerrero2013}
{Guerrero}, G., {Smolarkiewicz}, P.~K., {Kosovichev}, A.~G., \& {Mansour},
  N.~N. 2013, \apj, 779, 176

\bibitem[{{Jester} {et~al.}(2005){Jester}, {Schneider}, {Richards}, {Green},
  {Schmidt}, {Hall}, {Strauss}, {Vanden Berk}, {Stoughton}, {Gunn},
  {Brinkmann}, {Kent}, {Smith}, {Tucker}, \& {Yanny}}]{Jester2005}
{Jester}, S., {Schneider}, D.~P., {Richards}, G.~T., {et~al.} 2005, \aj, 130,
  873

\bibitem[{{K{\"a}pyl{\"a}} {et~al.}(2014){K{\"a}pyl{\"a}}, {K{\"a}pyl{\"a}}, \&
  {Brandenburg}}]{Kaepylae2014}
{K{\"a}pyl{\"a}}, P.~J., {K{\"a}pyl{\"a}}, M.~J., \& {Brandenburg}, A. 2014,
  \aap, 570, A43

\bibitem[{{Karak} {et~al.}(2014){Karak}, {K{\"a}pyl{\"a}}, {K{\"a}pyl{\"a}}, \&
  {Brandenburg}}]{Karak2014}
{Karak}, B.~B., {K{\"a}pyl{\"a}}, P.~J., {K{\"a}pyl{\"a}}, M.~J., \&
  {Brandenburg}, A. 2014, arXiv:1407.0984

\bibitem[{{K\H{o}v{\'a}ri} {et~al.}(2007){K\H{o}v{\'a}ri}, {Bartus},
  {Strassmeier}, {Vida}, {{\v S}vanda}, \& {Ol{\'a}h}}]{Kovari2007}
{K\H{o}v{\'a}ri}, Z., {Bartus}, J., {Strassmeier}, K.~G., {et~al.} 2007, \aap,
  474, 165

\bibitem[{{Kitchatinov} \& {R{\"u}diger}(2004)}]{Kitchatinov2004}
{Kitchatinov}, L.~L. \& {R{\"u}diger}, G. 2004, Astronomische Nachrichten, 325,
  496

\bibitem[{{Kovari} {et~al.}(2014){Kovari}, {Kriskovics}, {K{\"u}nstler},
  {Carroll}, {Strassmeier}, {Vida}, {Olah}, {Bartus}, \& {Weber}}]{Kovari2014}
{Kovari}, Z., {Kriskovics}, L., {K{\"u}nstler}, A., {et~al.} 2014, ArXiv
  e-prints

\bibitem[{{Lanza} {et~al.}(2014){Lanza}, {Das Chagas}, \& {De
  Medeiros}}]{Lanza2014}
{Lanza}, A.~F., {Das Chagas}, M.~L., \& {De Medeiros}, J.~R. 2014, \aap, 564,
  A50

\bibitem[{{Lund} {et~al.}(2014){Lund}, {Miesch}, \&
  {Christensen-Dalsgaard}}]{Lund2014}
{Lund}, M.~N., {Miesch}, M.~S., \& {Christensen-Dalsgaard}, J. 2014, \apj, 790,
  121

\bibitem[{{Reiners} \& {Schmitt}(2002)}]{Reiners2002a}
{Reiners}, A. \& {Schmitt}, J.~H.~M.~M. 2002, \aap, 384, 155

\bibitem[{{Reinhold} \& {Reiners}(2013)}]{Reinhold2013a}
{Reinhold}, T. \& {Reiners}, A. 2013, \aap, 557, A11

\bibitem[{{Reinhold} {et~al.}(2013){Reinhold}, {Reiners}, \&
  {Basri}}]{Reinhold2013b}
{Reinhold}, T., {Reiners}, A., \& {Basri}, G. 2013, \aap, 560, A4

\bibitem[{{Strassmeier} {et~al.}(2003){Strassmeier}, {Kratzwald}, \&
  {Weber}}]{Strassmeier2003}
{Strassmeier}, K.~G., {Kratzwald}, L., \& {Weber}, M. 2003, \aap, 408, 1103

\bibitem[{{Tkachenko} {et~al.}(2013){Tkachenko}, {Aerts}, {Yakushechkin},
  {Debosscher}, {Degroote}, {Bloemen}, {P{\'a}pics}, {de Vries}, {Lombaert},
  {Hrudkova}, {Fr{\'e}mat}, {Raskin}, \& {Van Winckel}}]{Tkachenko2013}
{Tkachenko}, A., {Aerts}, C., {Yakushechkin}, A., {et~al.} 2013, \aap, 556, A52

\bibitem[{{Uytterhoeven} {et~al.}(2011){Uytterhoeven}, {Moya},
  {Grigahc{\`e}ne}, {Guzik}, {Guti{\'e}rrez-Soto}, {Smalley}, {Handler},
  {Balona}, {Niemczura}, {Fox Machado}, {Benatti}, {Chapellier}, {Tkachenko},
  {Szab{\'o}}, {Su{\'a}rez}, {Ripepi}, {Pascual}, {Mathias},
  {Mart{\'{\i}}n-Ru{\'{\i}}z}, {Lehmann}, {Jackiewicz}, {Hekker},
  {Gruberbauer}, {Garc{\'{\i}}a}, {Dumusque}, {D{\'{\i}}az-Fraile}, {Bradley},
  {Antoci}, {Roth}, {Leroy}, {Murphy}, {De Cat}, {Cuypers}, {Kjeldsen},
  {Christensen-Dalsgaard}, {Breger}, {Pigulski}, {Kiss}, {Still}, {Thompson},
  \& {van Cleve}}]{Uytterhoeven2011}
{Uytterhoeven}, K., {Moya}, A., {Grigahc{\`e}ne}, A., {et~al.} 2011, \aap, 534,
  A125

\bibitem[{{Vida} {et~al.}(2007){Vida}, {K\H{o}v{\'a}ri}, {{\v S}vanda},
  {Ol{\'a}h}, {Strassmeier}, \& {Bartus}}]{Vida2007}
{Vida}, K., {K\H{o}v{\'a}ri}, Z., {{\v S}vanda}, M., {et~al.} 2007,
  Astronomische Nachrichten, 328, 1078

\bibitem[{{Vida} {et~al.}(2014){Vida}, {Ol{\'a}h}, \& {Szab{\'o}}}]{Vida2014}
{Vida}, K., {Ol{\'a}h}, K., \& {Szab{\'o}}, R. 2014, \mnras, 441, 2744

\bibitem[{{Weber} {et~al.}(2005){Weber}, {Strassmeier}, \&
  {Washuettl}}]{Weber2005}
{Weber}, M., {Strassmeier}, K.~G., \& {Washuettl}, A. 2005, Astronomische
  Nachrichten, 326, 287

\bibitem[{{Zechmeister} \& {K{\"u}rster}(2009)}]{Zechmeister2009}
{Zechmeister}, M. \& {K{\"u}rster}, M. 2009, \aap, 496, 577

\end{thebibliography}
\bibliographystyle{aa}

\end{document}